\documentclass[twocolumn]{aa}
\usepackage{graphicx}
\usepackage{txfonts}
\def\etal{{et al. \rm}}
\def\for{[\ion{O}{i}] $\lambda$6300}
\def\tri{\ion{O}{i} triplet}
\def\fors{[\ion{O}{i}] $\lambda$6300 }
\def\tris{\ion{O}{i} triplet }
\def\ca{\ion{Ca}{ii} H+K}
\def\cas{\ion{Ca}{ii} H+K }
\begin{document}
   \title{On the determination of oxygen abundances in chromospherically active stars\thanks{Based on observations collected at the European Southern Observatory, Chile (Proposals 64.L-0249 and 071.D-0260).}\fnmsep\thanks{Table~\ref{tab_data} is only available in electronic form at the CDS via anonymous ftp to {\tt cdsarc.u-strasbg.fr (130.79.128.5)} or via {\tt http://cdsweb.u-strasbg.fr/cgi-bin/qcat?]/A+A/vol/page}}}
   \titlerunning{Oxygen abundances in chromospherically active stars}

   \author{T. Morel
          \and
          G. Micela}

   \offprints{T. Morel,
              \email{morel@astropa.unipa.it}}

   \institute{Istituto Nazionale di Astrofisica, Osservatorio Astronomico di Palermo G.\,S. Vaiana, Piazza del Parlamento 1, I-90134 Palermo, Italy}

   \date{Received 2 March 2004; accepted 29 April 2004}

   \abstract{We discuss oxygen abundances derived from \fors and the \tris in stars spanning a wide range in chromospheric activity level, and show that these two indicators yield increasingly discrepant results with higher chromospheric/coronal activity measures. While the forbidden and permitted lines give fairly consistent results for solar-type disk dwarfs, spuriously high \tris abundances are observed in young Hyades and Pleiades stars, as well as in individual components of RS CVn binaries (up to 1.8 dex). The distinct behaviour of the [\ion{O}{i}]-based abundances which consistently remain near-solar suggests that this phenomenon mostly results from large departures from LTE affecting the \tris at high activity level that are currently unaccounted for, but also possibly from a failure to adequately model the atmospheres of K-type stars. These results suggest that some caution should be exercised when interpreting oxygen abundances in active binaries or young open cluster stars.

   }

   \maketitle
%
%________________________________________________________________

\section{Introduction} \label{sect_intro}
The determination of oxygen abundances in distinct stellar populations (bulge, thin/thick disk, and halo) is of fundamental importance for a better understanding of the chemical evolution of the Galaxy and its formation history (e.g., McWilliam 1997). On the other hand, oxygen is a key ingredient in evolution codes, as it affects the energy generation and opacity in stellar interiors. This is of particular relevance for issues related to the lithium depletion in open cluster stars, for instance (e.g., Piau \& Turck-Chi\`eze 2002). 

For practical reasons, oxygen abundances have been traditionally determined from transitions in the optical domain, either the 7774 \AA-\tris or the forbidden \fors and [\ion{O}{i}] $\lambda$6363 lines (these spectral diagnostics can be complemented by molecular UV and IR OH bands: e.g., Boesgaard \etal 1999). Unfortunately, the former is bound to be affected by significant, yet poorly-constrained non-LTE (NLTE) effects. Although the forbidden lines are immune to departures from LTE, they are generally very weak and become virtually immeasurable in metal-poor dwarfs, which are potentially among the most interesting targets. Additionally, they also suffer from blending with weak CN or Ni lines (see below). Both indicators are very sensitive to the effective temperature and, to a lesser extent, to the surface gravity assumed. Furthermore, the treatment of granulation may also be important (Asplund \etal 2004). The long-standing discrepancy often observed between these two indicators (the permitted lines generally yielding systematically higher abundances than the forbidden transitions) likely stems from these caveats, although the cause of this disparity is still, despite much work, not well understood (e.g., Fulbright \& Johnson 2003). 

During the course of our abundance study of a sample of single-lined RS CVn binaries (Morel \etal 2003; hereafter M03, Morel \etal 2004; hereafter M04), we noticed a dramatic discrepancy between the oxygen abundances given by \fors and the \tri, with the latter yielding suspiciously high values. Using these observations, supplemented by data for Pleiades, Hyades, and field FG dwarfs, here we show that the [\ion{O}{i}]- and \ion{O}{i}-based abundances diverge with increasing stellar activity level.

\section{Observational data} \label{sect_results}
Spectra of 14 single-lined chromospherically active binaries from the list of Strassmeier et al. (1993) were obtained in 2000 and 2003 at the ESO 1.5/2.2-m telescopes at La Silla (Chile) with the echelle spectrograph FEROS (resolving power $R$$\sim$48\,000). The spectral range covered is 3600--9200 \AA, hence enabling a {\em simultaneous} coverage of \ca, \for, and the \tris (the extremely weak [\ion{O}{i}] $\lambda$6363 line could not be reliably measured in our spectra). We also observed a control sample made up of 7 stars with similar characteristics (i.e., K2--G8 subgiants), but with low X-ray luminosities (H\"unsch, Schmitt, \& Voges 1998a). 

We refer the reader to M03 and M04 for details on the reduction procedure and abundance analysis. Briefly, the effective temperature and surface gravity were derived from the excitation and ionization equilibrium of the iron lines, while the microturbulent velocity was determined by requiring the \ion{Fe}{i} abundances to be independent of the line strength. The oxygen abundances were derived from a differential analysis using plane-parallel, line-blanketed LTE Kurucz atmospheric models (with a length of the convective cell over the pressure scale height, $\alpha$=$l/H_{\rm p}$=0.5, and no overshoot) and the current version of the MOOG software. The choice of another 1-D model (e.g., MARCS) is very unlikely to substantially affect the conclusions presented in the following (see Fulbright \& Johnson 2003). All $gf$-values were calibrated with a high S/N moonlight spectrum acquired with the same instrumental setup. The values adopted for the oxygen features and the measured equivalent widths (EWs) are given in Table~\ref{tab_ew}. The EWs of \fors have been corrected for the contribution of the high-excitation \ion{Ni}{i} $\lambda$6300.339 line ($\chi$=4.27 eV). The corresponding EW was estimated by a curve-of-growth analysis using the derived atmospheric parameters and Ni abundances (M03, M04), and assuming $\log gf$=--2.31 (Allende Prieto, Lambert, \& Asplund 2001). In order to be consistent with data for Pleiades stars (Schuler \etal 2004), we use this value instead of a recent (and presumably more accurate) laboratory determination ($\log gf$=--2.11; Johansson \etal 2003). Albeit extremely weak (typically EW$\sim$3--4 m\AA), the Ni line may contribute up to 15\% to the total EW. A similar procedure was used to estimate the oscillator strength of \fors from the solar spectrum. 

\begin{table}
\centering
\caption{Oscillator strengths, first excitation potentials, and EWs of the oxygen features (in m\AA). A blank indicates that the EW was not reliably measurable because of an unsatisfactory Gaussian fit or cosmic rays/telluric features affecting the line profile. The latter issue is particularly relevant in the case of \fors (we used the telluric atlas of Hinkle \etal 2000).}
\label{tab_ew}
%\hspace*{0.4cm}
%\scriptsize
\begin{tabular}{lcccc} \hline\hline
$\lambda$ (\AA) &  6300.304$^a$ & 7771.944 & 7774.166 & 7775.388\\\hline
$\log gf$ & --9.778 & 0.297 & 0.114 & --0.064\\
$\chi$ (eV) & 0.000  & 9.147 & 9.147 & 9.147\\\hline
\object{HD 28    } &      & 35.8 & 36.5 & 32.0 \\
\object{HD 1227  } &      & 58.1 & 52.7 & \\
\object{HD 4482  } & 25.1 & 68.2 &      & 47.3\\
\object{HD 10909 } & 18.7 & 65.3 & 54.4 & 38.7\\
\object{HD 17006 } &      & 64.4 &      & \\
\object{HD 19754 } & 21.8 &      & 62.5 & 55.7\\
\object{HD 72688 } &      & 81.6 &      & 58.3\\
\object{HD 83442 } &      & 84.0 & 75.8 &\\
\object{HD 113816} & 20.8 & 92.3 &      & \\
\object{HD 118238} &      & 125.1& 120.7&\\
\object{HD 119285} &      & 93.8 & 84.2 & 53.9\\
\object{HD 154619} & 19.2 & 62.9 & 59.2 & 45.6\\
\object{HD 156266} & 34.3 &      &      & \\
\object{HD 181809} & 13.4 & 64.5 &      & 44.8\\
\object{HD 182776} &      &      & 95.7 & \\
\object{HD 202134} & 22.7 & 86.3 &      & \\
\object{HD 204128} &      & 82.1 &      & \\
\object{HD 205249} &      & 95.4 & 92.4 & 64.1\\
\object{HD 211391} & 27.7 & 60.0 &      & 44.3\\
\object{HD 217188} &      & 76.8 &      & 52.7\\
\object{HD 218527} &      & 52.2 & 49.8 & \\\hline
\end{tabular}
\begin{flushleft}
\normalsize $^a$ Corrected for the contribution of \ion{Ni}{i} $\lambda$6300.339 (see text).
\end{flushleft}
\end{table}

The bulk of the oxygen abundances for Pleiades stars come from Schuler \etal (2004), with the exception of HII 676 (King \etal 2000). For Hyades stars, we use the data of Garc\'{\i}a L{\'o}pez \etal (1993) and King \& Hiltgen (1996). For the nearby main sequence FG stars, we restrict ourselves to objects with measurements in both \fors and the \tris (Bensby, Feltzing, \& Lundstr\"om 2004; King \& Boesgaard 1995; Reddy \etal 2003). Among these sources, only King \& Boesgaard (1995) and King \& Hiltgren (1996) did not take into account the \ion{Ni}{i} $\lambda$6300.3 line. Their \fors abundances are thus likely to be slightly overestimated. In order to avoid oxygen overabundances arising from contamination from Type II supernova ejecta, care has been taken to only select stars with solar metallicities and/or kinematical properties typical of the thin disk population. The oxygen abundances from the literature have been rescaled to our adopted oxygen and iron solar abundances. In order to be consistent with Kurucz models and opacities, we assumed $\log$ $\epsilon_{\odot}$(O)=8.93 and $\log$ $\epsilon_{\odot}$(Fe)=7.67. In addition, {\em all} abundances given by the \tris have been uniformly converted into NLTE values using the corrections of Gratton \etal (1999). The abundances are systematically revised downwards by a small amount never exceeding 0.13 dex.       

\begin{figure*}
\resizebox{\hsize}{!}
{\rotatebox{0}{\includegraphics{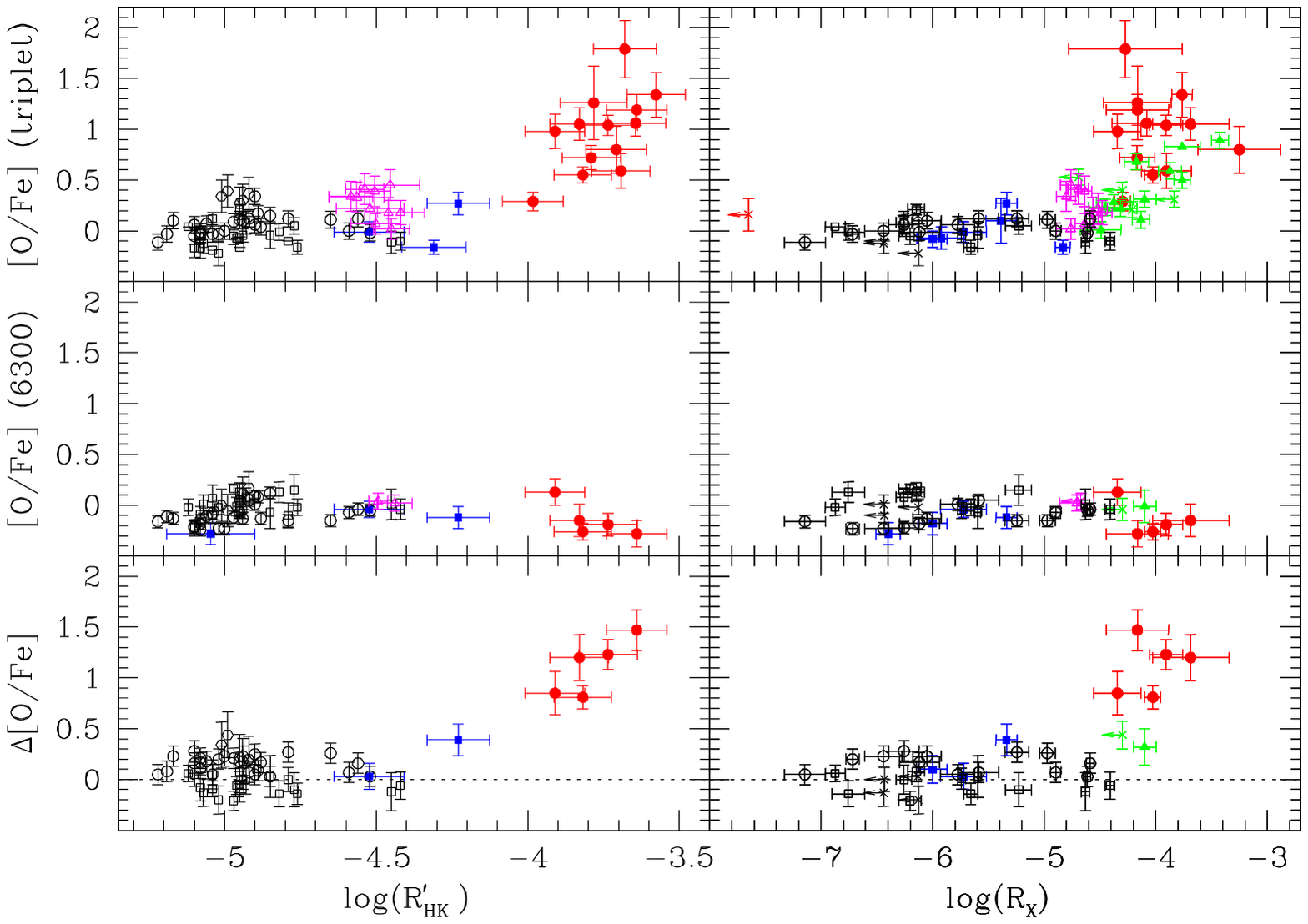}}}
\caption{Oxygen abundances as a function of the activity indices $R_{\rm HK}^{\prime}$ and $R_{\rm X}$. The bottom panels show the difference between [O/Fe] given by the \tris and the \fors line. {\em Filled circles}: RS CVn binaries (M03; M04), {\em filled squares}: field subgiants (M04), {\em filled triangles}: Pleiades stars (King \etal 2000; Schuler \etal 2004), {\em open triangles}: Hyades stars (Garc\'{\i}a L{\'o}pez \etal 1993; King \& Hiltgen 1996), {\em open circles, squares, and hexagons}: disk dwarfs (from Bensby \etal 2004, King \& Boesgaard 1995, and Reddy \etal 2003, respectively). The colour coding in the online version of this journal is the following: {\em red}: RS CVn binaries, {\em blue}: field subgiants, {\em green}: Pleiades stars, {\em magenta}: Hyades stars.  The crosses in the right-hand panels are upper limits.} 
\label{fig1}
\end{figure*}

The activity index, $R_{\rm HK}^{\prime}$, defined as the radiative loss in the \cas lines in units of the bolometric luminosity (after correction for photospheric contribution), was used as a primary indicator of chromospheric activity. This quantity was derived from our spectra following Linsky et al. (1979). In view of the $(V-R)$ and $(V-I)$ colour anomalies exhibited by active stars (see M04), we did not use the observed $(V-R)$ value in the derivation of the absolute flux in the 3925--3975 \AA \ wavelength range, ${\cal F}_{50}$. Instead, we computed the colour appropriate for a star with the derived effective temperature and iron abundance (eq. [5] of Alonso, Arribas, \& Mart\'{\i}nez-Roger 1999). The bolometric luminosities were estimated from theoretical isochrones. The $R_{\rm HK}^{\prime}$ data for disk dwarfs were gathered from the literature (Soderblom 1985; Henry \etal 1996; Wright \etal 2004). The data for Hyades stars come from Paulson \etal (2002). For 8 stars in this sample, there is a good agreement with previous measurements (Duncan \etal 1984). We also define $R_{\rm X}$, which is given as the ratio between the X-ray (in the {\em ROSAT} [0.1--2.4 keV] energy band) and bolometric luminosities. More details regarding the derivation of $R_{\rm HK}^{\prime}$ and $R_{\rm X}$ can be found in M03. The oxygen abundances and activity indices, as well as the source of the data, are given in Table~\ref{tab_data} (only available in electronic form).

\section{Results} \label{sect_results}
Figure~\ref{fig1} shows the oxygen abundances yielded by \fors and the \tri, as a function of the activity indices $R_{\rm HK}^{\prime}$ and $R_{\rm X}$. As can be seen, the abundances given by both set of lines are in fair agreement for solar-type field stars. However, there is a steeply increasing discrepancy as one goes to higher activity levels. In sharp contrast to the behaviour of the \tri, the \for-based abundances remain roughly solar even for the most active stars. There is a hint that the Pleiades stars exhibit lower triplet abundances than the active binaries at a given $R_{\rm X}$ value. Several inconsistencies (e.g., temperature scale) are likely to introduce systematics when comparing abundances taken from various works in the literature. However, such an offset might arise from a temperature effect (see Sect.~\ref{sect_discussion}). We note that the choice of the atomic data is of little relevance here. The oscillator strengths adopted in the various literature sources (including our values) lie within 0.08 dex, i.e., a value translating into abundance  differences typically comparable to the uncertainties. 

We cannot rule out the existence of a systematic underestimation of the iron abundances (at the $\sim$0.1--0.2 dex level) in the RS CVn binaries (M04). Although this would evidently significantly affect the [O/Fe] values, the difference between the abundances derived from the permitted and forbidden lines is independent of the iron content and is thus a robust indicator of the influence of chromospheric activity. 

The correlation between the \ion{O}{i}-based abundances and $R_{\rm X}$ is remarkably tight considering the following points.\footnote{In the Pleiades dataset alone, [O/Fe] (triplet) is correlated with $R_{\rm X}$  at a confidence level exceeding 98\%, whether or not upper limits are considered (we made us of the Kendall's $\tau$ method in the case of censored data; Isobe, Feigelson \& Nelson 1986).} Firstly, we recall that the measurements of the oxygen abundances and X-ray luminosities are not, contrary to the \cas data, co-temporal. In addition to flare-like events, the X-ray emission in RS CVn binaries and young open cluster stars can be intrinsically variable on long timescales (e.g., Kashyap \& Drake 1999; Micela \etal 1996). Secondly, in contrast to the X-ray data which diagnose the coronal regions, the \cas lines are spectral diagnostics of the lower chromosphere and are as such better probes of the physical conditions prevailing in vicinity of the stellar photosphere. 

\section{Discussion} \label{sect_discussion}
As mentioned previously, the [\ion{O}{i}] and \ion{O}{i} lines are very sensitive to the choice of the atmospheric model. It is therefore conceivable that the trend observed in Fig.~\ref{fig1} is actually an artefact of systematics in the determination of the atmospheric parameters. In the case of the active binaries, our excitation temperatures appear marginally higher than values derived from $T_{\rm eff}$-($B-V$) calibrations based on the infrared flux method (Alonso \etal 1999): $<$$T_{\rm exc}$--$T_{\rm colour}$$>$=+80$\pm$46 K (M04). On the other hand, our surface gravities derived from the ionization equilibrium of the iron lines tend to be lower than values derived from theoretical isochrones ($<$$\log g_{\rm ioni}$--$\log g_{\rm iso}$$>$=--0.21$\pm$0.06 dex). Such zero-point offsets may have various causes (e.g., NLTE effects, identification of model label temperatures with values derived from integrated flux). Regardless, we examined the potential impact of systematic errors of this magnitude on the resulting oxygen abundances, as well as the effect of the treatment of convective energy transport (overshoot option switched on/off), as it modifies the $T$($\tau$) relation in the deepest photospheric layers where the \tris is formed. Tests were also performed with models characterized by different opacities (by means of an enrichment in $\alpha$ elements). Since the excitation and ionization equilibrium of the iron lines must be simultaneously satisfied, note that changes in $T_{\rm eff}$ are necessarily accompanied by variations in $\log g$ and {\em vice versa}. As expected, errors in $T_{\rm eff}$ and $\log g$ significantly bias the [O/Fe] determinations (see Table~\ref{tab_par}). However, resolving the discrepancy between the oxygen indicators would require alterations of these quantities well beyond any reasonable uncertainty. More importantly, the differences between the excitation/$(B-V)$-colour temperatures and the ionization/isochrone gravities are not correlated with the activity indices. This rules out the possibility that the activity trend results from an increasingly underestimated temperature at high activity levels, for instance. The presence of a faint, cooler companion is also not thought to significantly affect our $T_{\rm eff}$ determinations (Katz \etal 2003).

\addtocounter{table}{+1}
\begin{table}
\centering
\caption{Effect of changes in the stellar parameters of \object{HD 202134} on the oxygen abundances. $\Delta$[O/Fe] is defined as [O/Fe] (triplet)--[O/Fe] (6300).}
\label{tab_par}
%\hspace*{0.4cm}
%\scriptsize
\begin{tabular}{lcccc} \hline\hline
                 & [Fe/H] & [O/Fe] & [O/Fe] & $\Delta$[O/Fe]\\
                 &        & 6300 & triplet & \\\hline
$\Delta T_{\rm eff}$=+150 K     &  +0.08 &  +0.14 & --0.18 & --0.32\\
$\Delta \log g$=+0.25 dex [cgs] &  +0.04 &  +0.08 & --0.10 & --0.18\\
$\Delta \xi$=+0.20 km s$^{-1}$   & --0.08 &  +0.08 &  +0.01 & --0.07\\
$\Delta$[$\alpha$/Fe]=--0.2             & --0.01 & --0.01 &  +0.01 &  +0.02\\
with overshooting               &   +0.00 &  +0.01 &  +0.01 &  +0.00\\\hline
\end{tabular}
\end{table}

\begin{figure}
\resizebox{\hsize}{!}
{\rotatebox{0}{\includegraphics{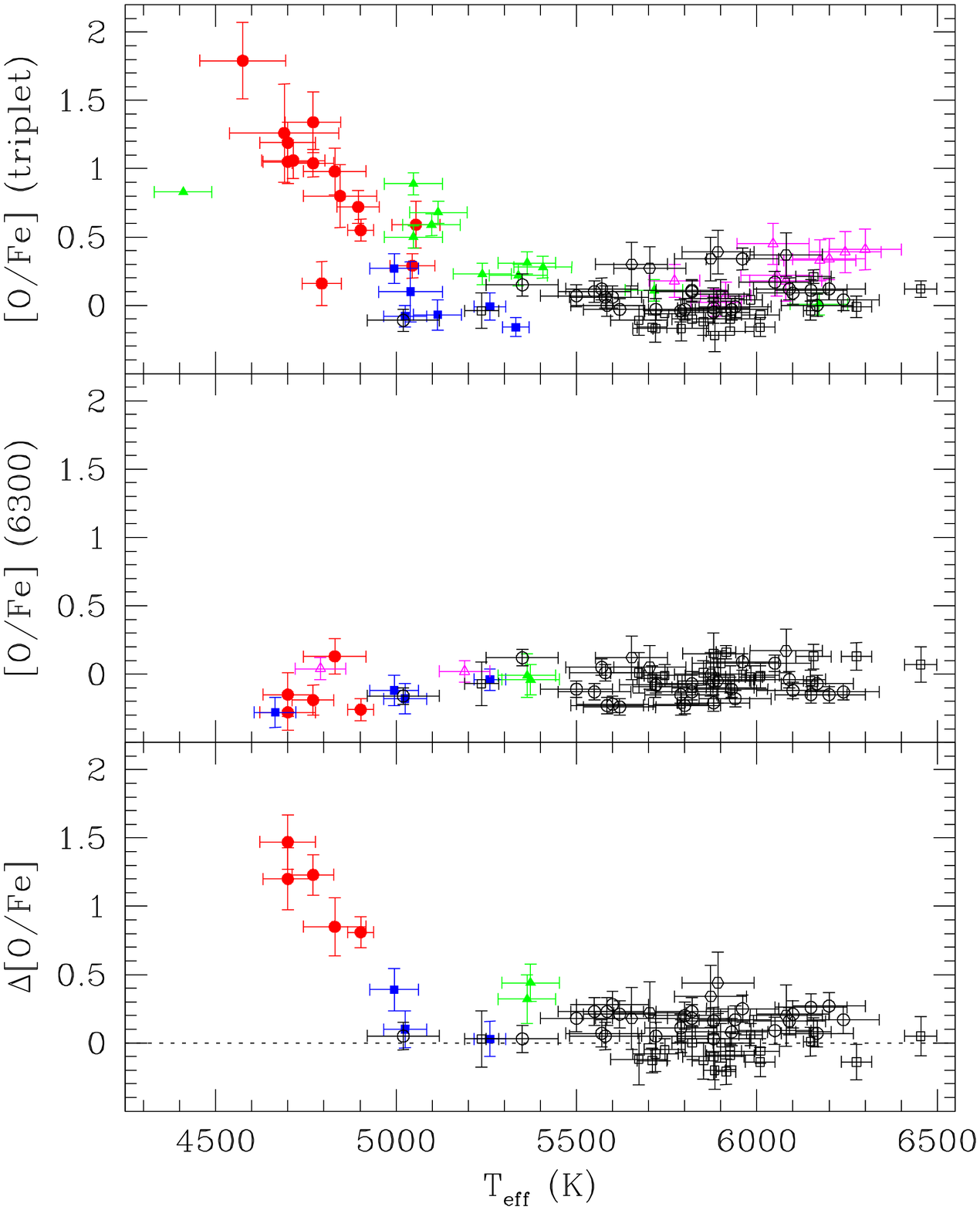}}}
\caption{Oxygen abundances as a function of the effective temperature. The bottom panel shows the difference between [O/Fe] given by the \tris and the \fors line. Symbols as in Fig~\ref{fig1}.}
\label{fig2}
\end{figure}

Let us now examine the potential importance of physical processes causally related to activity. Following Neff, O'Neal, \& Saar (1995), we investigated the effect of photospheric spots by carrying out an abundance analysis on a composite, synthetic Kurucz spectrum of a fictitious star with $T_{\rm eff}$=4830 K covered by cooler regions ($T_{\rm eff}$=3830 K) with different covering factors, $f_s$ (see M03). Although the effect is appreciable, we find that the oxygen discrepancy only increases by 0.08 and 0.12 dex for $f_s$=30 and 50\%, respectively. We assessed in M03 the importance of chromospheric heating in inducing abundance anomalies by altering the temperature structure of the Kurucz models: a temperature gradient reversal in the outermost regions was incorporated, as well as a heating (up to 190 K) of the deeper photospheric layers (see fig.3 of M03). We considered two model chromospheres meant to be representative of K-type giants (Kelch et al. 1978) and RS CVn binaries (Lanzafame, Bus\`a, \& Rodon\`o 2000). A LTE abundance analysis was then performed on \object{HD 10909} using a classical Kurucz model and these two sets of model atmospheres. To fulfil the excitation and ionization equilibrium of the iron lines, it was necessary in the two latter cases to use an 'underlying' Kurucz model with a much lower effective temperature and gravity. Here we adopt a different approach by keeping the parameters of the atmospheric model unchanged (only the microturbulence was adjusted). Because of their inverse sensitivity to temperature changes, it comes as no surprise that merging a chromospheric component on the Kurucz model leads to a better agreement between the oxygen indicators. However, the discrepancy is 'only' reduced by 0.17 and 0.28 dex for the model chromospheres of Kelch et al. (1978) and Lanzafame \etal (2000), respectively. This seems insufficient to account for the observations. 

Apart from a global temperature rise, however, it is likely that a chromosphere would primarily manifest itself by overionization/overexcitation effects. In the case of the Sun, for instance, the solar chromosphere must be taken into account when studying the NLTE line formation of the \tris and, in particular, to reproduce the centre-to-limb behaviour of these features (Takeda 1995). To our knowledge, all comprehensive NLTE line formation studies published to date are based on conventional atmospheric models without a chromospheric component. Several investigations have concentrated on FG dwarfs, but considerably less attention has been unfortunately paid to cooler stars. The NLTE corrections of Gratton \etal (1999) can be very large for warm, low gravity stars (up to 1 dex), but are modest for K-type subgiants (less than 0.13 dex for the stars in our sample; similar results were obtained by Takeda 2003). When applied to our abundances, they only slightly reduce the disparity between the oxygen indicators. Yong \etal (2004) recently evidenced dramatic overionization effects in Pleiades dwarfs, with an increasing discrepancy between the \ion{Fe}{i}- and \ion{Fe}{ii}-based abundances with decreasing $T_{\rm eff}$. This highlights the unexpectedly large magnitude of the NLTE effects in cool, active stars. Since the \tris is known to be liable to large departures from LTE, it is tempting to associate the oxygen abundance anomalies observed in Pleiades stars to this phenomenon (Schuler \etal 2004). The fact that \for, which is immune to these effects, consistently yields roughly near-solar abundances supports a similar interpretation in our sample (Fig.~\ref{fig1}). 

Tomkin \etal (1992) reported an increasing discrepancy between the oxygen indicators with decreasing effective temperatures in halo main sequence FG stars. In a similar vein, Schuler \etal (2004) found that cool stars in the Pleiades exhibit higher \ion{O}{i}-based oxygen abundances. This may primarily be an activity effect in view of the tight correlation between ($B-V)$$_0$ and $R_{\rm X}$ (Micela \etal 1999). There is also some evidence in our data for a relationship between $T_{\rm eff}$ and [O/Fe] (Fig.~\ref{fig2}). This might arise from the fact that the most active stars (i.e., the RS CVn binaries) turn out to be the coolest objects in our sample. It is worth noting that a similar relation also seems to hold for presumably inactive, metal-poor (sub)giants (see fig.10 of Fulbright \& Johnson 2003), although the difference is much more marked here ($\Delta$[O/Fe]=1.8 against 0.5 dex at $T_{\rm eff}$$\sim$4700 K). If confirmed at near-solar metallicity, this trend would suggest that the same mechanism acts in both samples of active and inactive stars to yield spuriously high abundances from the \tri. 

To summarize, we cannot rule out either of these two (non mutually exclusive) possibilities: (a) the discrepancy between the oxygen indicators results from inadequacies in the atmospheric models for cool stars (e.g., inappropriate temperature structure in the deepest layers; Carretta, Gratton, \& Sneden 2000; Israelian \etal 2004), this problem becoming more acute for chromospherically active stars, or (b) NLTE effects are dramatically underestimated in cool, active stars and, to a lesser extent, in their inactive analogues. 

\section{Summary} \label{sect_summary}
As discussed in this paper, \fors appears relatively free of the problems affecting the \tri. This makes it a much better oxygen abundance indicator in stars with a very high level of chromospheric activity, such as members of young open clusters or active, tidally-locked binaries. Although using molecular OH bands may constitute an alternative, having to solely rely on the \fors line appears rather unfortunate considering the weakness of this feature in open cluster dwarfs, for instance (EW $\la$ 10 m\AA). On a more optimistic note, our observations have unveiled a missing ingredient in our modelling of the \tris in active, cool stars. Although challenging, more progress on this issue can be expected from the theoretical side: more realistic models for K-type stars (e.g., incorporating thermal inhomogeneities; Nissen \etal 2002) and detailed NLTE calculations applied to chromospherically active stars.       

\begin{acknowledgements}
  This research was supported through a European Community Marie Curie
  Fellowship (No. HPMD-CT-2000-00013). G.\,M. acknowledges financial support from MIUR (Ministero della Istruzione,
  dell'Universit\`a e della Ricerca). We wish to thank Dr. S. C. Schuler for making his manuscript available to us prior to publication and an anonymous referee for useful comments. This research made use of NASA's Astrophysics Data System Bibliographic Services and the Simbad database operated at CDS, Strasbourg, France.  
\end{acknowledgements}

\addtocounter{table}{-2}
\begin{table*}
\centering
\caption{Oxygen abundances and activity indices. The [O/Fe] data from the literature were rescaled to our adopted iron and oxygen solar abundances ($\log$ $\epsilon_{\odot}$[Fe]=7.67 and $\log$ $\epsilon_{\odot}$[O]=8.93), and corrected for departures from LTE in the case of the \tris (Gratton \etal 1999). We use the values of Schuler \etal (2004) derived without convective overshoot ('NOVER').}
\label{tab_data}
%\hspace*{0.4cm}
\scriptsize
\begin{tabular}{lcrrrcccccc} \hline\hline
ID$^a$    &    $T_{\rm eff}$ & \multicolumn{1}{c}{[Fe/H]$^b$}   &     \multicolumn{1}{c}{[O/Fe] (6300)}  &     \multicolumn{1}{c}{[O/Fe] (Triplet)}
 & Ref. & log($R_{\rm HK}^{\prime}$)$^c$ & Ref. & log($L_{\rm X}$)$^d$ &  log($R_{\rm X}$)$^d$  &  Ref.\\
   &  (K) &  &  &  &  & &  & (ergs s$^{-1}$) &    &  \\\hline
\multicolumn{2}{l}{\bf RS CVn binaries} &&&&&&&&&\\
\object{HD     28} & 4794 &    0.01$\pm$0.08     &                  &    0.16$\pm$0.16 & 1 &                 &     & $<$27.34       & $<$--7.65 & 14 \\
\object{HD  10909} & 4830 &  --0.41$\pm$0.09     &    0.13$\pm$0.13 &    0.98$\pm$0.17 & 2 & --3.91$\pm$0.10 & 2   & 30.26$\pm$0.21 & --4.34$\pm$0.21 & 15\\
\object{HD  19754} & 4700 &  --0.39$\pm$0.08     &  --0.15$\pm$0.16 &    1.05$\pm$0.16 & 1 & --3.83$\pm$0.10 & 1   & 31.43$\pm$0.34 & --3.69$\pm$0.34 & 16\\
\object{HD  72688} & 5045 &    0.11$\pm$0.07     &                  &    0.29$\pm$0.09 & 2 & --3.98$\pm$0.10 & 2   & 30.96$\pm$0.07 & --4.30$\pm$0.07 & 15\\
\object{HD  83442} & 4715 &    0.02$\pm$0.10     &                  &    1.06$\pm$0.13 & 2 & --3.64$\pm$0.10 & 2   & 30.87$\pm$0.31 & --4.08$\pm$0.31 & 14 \\
\object{HD 113816} & 4700 &  --0.11$\pm$0.09     &  --0.28$\pm$0.13 &    1.19$\pm$0.15 & 2 & --3.64$\pm$0.10 & 2   & 31.09$\pm$0.28 & --4.16$\pm$0.28 & 14 \\ 
\object{HD 118238} & 4575 &  --0.12$\pm$0.13     &                  &    1.79$\pm$0.28 & 2 & --3.68$\pm$0.10 & 2   & 31.66$\pm$0.51 & --4.27$\pm$0.51 & 15\\
\object{HD 119285} & 4770 &  --0.23$\pm$0.10     &                  &    1.34$\pm$0.22 & 2 & --3.58$\pm$0.10 & 2   & 30.56$\pm$0.09 & --3.77$\pm$0.09 & 14 \\
\object{HD 181809} & 4902 &  --0.09$\pm$0.04     &  --0.26$\pm$0.08 &    0.55$\pm$0.08 & 1 & --3.82$\pm$0.10 & 1   & 30.92$\pm$0.07 & --4.03$\pm$0.07 & 14 \\
\object{HD 182776} & 4690 &  --0.06$\pm$0.16     &                  &    1.26$\pm$0.36 & 1 & --3.78$\pm$0.11 & 1   & 30.88$\pm$0.30 & --4.16$\pm$0.30 & 17\\
\object{HD 202134} & 4770 &  --0.06$\pm$0.07     &  --0.19$\pm$0.11 &    1.04$\pm$0.10 & 1 & --3.74$\pm$0.10 & 1   & 30.99$\pm$0.15 & --3.91$\pm$0.15 & 15\\
\object{HD 204128} & 4845 &  --0.02$\pm$0.11     &                  &    0.80$\pm$0.23 & 1 & --3.71$\pm$0.10 & 1   & 31.50$\pm$0.37 & --3.25$\pm$0.37 & 15\\
\object{HD 205249} & 5055 &    0.12$\pm$0.09     &                  &    0.59$\pm$0.17 & 1 & --3.69$\pm$0.10 & 1   & 31.27$\pm$0.22 & --3.90$\pm$0.22 & 16\\
\object{HD 217188} & 4895 &  --0.16$\pm$0.07     &                  &    0.72$\pm$0.12 & 1 & --3.79$\pm$0.10 & 1   & 30.69$\pm$0.16 & --4.17$\pm$0.16 & 16\\
\multicolumn{2}{l}{\bf Field subgiants} &&&&&&&&&\\
\object{HD   1227} & 5115 &    0.19$\pm$0.07     &                  &  --0.07$\pm$0.11 & 1 &                 &     & 29.39$\pm$0.23 & --5.92$\pm$0.23 & 18\\
\object{HD   4482} & 4995 &    0.05$\pm$0.07     &  --0.12$\pm$0.11 &    0.27$\pm$0.11 & 1 & --4.23$\pm$0.10 & 1   & 29.83$\pm$0.10 & --5.34$\pm$0.10 & 18\\
\object{HD  17006} & 5332 &    0.36$\pm$0.06     &                  &  --0.16$\pm$0.07 & 1 & --4.31$\pm$0.10 & 1   & 29.45$\pm$0.07 & --4.83$\pm$0.07 & 18\\
\object{HD 154619} & 5260 &    0.13$\pm$0.06     &  --0.04$\pm$0.08 &  --0.01$\pm$0.10 & 1 & --4.52$\pm$0.11 & 1   & 29.52$\pm$0.20 & --5.72$\pm$0.20 & 18\\
\object{HD 156266} & 4665 &    0.25$\pm$0.08     &  --0.28$\pm$0.11 &                  & 1 & --5.05$\pm$0.12 & 1   & 28.97$\pm$0.18 & --6.46$\pm$0.18 & 18\\
\object{HD 211391} & 5025 &    0.23$\pm$0.07     &  --0.18$\pm$0.11 &  --0.08$\pm$0.08 & 1 &                 &     & 29.42$\pm$0.13 & --6.00$\pm$0.13 & 18\\
\object{HD 218527} & 5040 &  --0.08$\pm$0.09     &                  &    0.10$\pm$0.22 & 1 &                 &     & 29.86$\pm$0.14 & --5.39$\pm$0.14 & 18\\
\multicolumn{2}{l}{\bf Pleiades stars} &&&&&&&&&\\
\object{HII   193} & 5339 &    0.06$\pm$0.05     &                  &    0.22$\pm$0.08 & 3 &                 &     & 28.97$\pm$0.15 & --4.31$\pm$0.15 & 19\\   
\object{HII   250} & 5715 &    0.06$\pm$0.05     &                  &    0.11$\pm$0.08 & 3 &                 &     & 29.39$\pm$0.08 & --4.13$\pm$0.08 & 20\\
\object{HII   263} & 5048 &    0.06$\pm$0.05     &                  &    0.89$\pm$0.08 & 3 &                 &     & 29.79$\pm$0.08 & --3.42$\pm$0.08 & 19\\
\object{HII   298} & 5048 &    0.06$\pm$0.05     &                  &    0.50$\pm$0.08 & 3 &                 &     & 29.72$\pm$0.07 & --3.76$\pm$0.07 & 19\\
\object{HII   571} & 5373 &    0.06$\pm$0.05     &  --0.04$\pm$0.11 &    0.40$\pm$0.08 & 3 &                 &     & $<$29.03       &   $<$--4.30     & 20\\
\object{HII   676} & 4410 &    0.06$\pm$0.05     &                  &    \multicolumn{1}{c}{0.83}          & 4 &                 &     & 29.00$\pm$0.16 & --3.76$\pm$0.16 & 20\\     
\object{HII   746} & 5407 &    0.06$\pm$0.05     &                  &    0.28$\pm$0.08 & 3 &                 &     & 28.91$\pm$0.15 & --4.38$\pm$0.15 & 19\\   
\object{HII   916} & 5098 &    0.06$\pm$0.05     &                  &    0.59$\pm$0.08 & 3 &                 &     & 29.27$\pm$0.10 & --3.87$\pm$0.10 & 19\\
\object{HII  1593} & 5407 &    0.06$\pm$0.05     &                  &    0.21$\pm$0.08 & 3 &                 &     & $<$29.14       &   $<$--4.21     & 21\\ 
\object{HII  2126} & 5142 &    0.06$\pm$0.05     &                  &    0.31$\pm$0.08 & 3 &                 &     & $<$29.33       &   $<$--3.83     & 21\\ 
\object{HII  2284} & 5363 &    0.06$\pm$0.05     &  --0.01$\pm$0.16 &    0.31$\pm$0.08 & 3 &                 &     & 29.16$\pm$0.10 & --4.10$\pm$0.10 & 19\\
\object{HII  2311} & 5239 &    0.06$\pm$0.05     &                  &    0.23$\pm$0.08 & 3 &                 &     & 28.82$\pm$0.15 & --4.45$\pm$0.15 & 19\\   
\object{HII  2462} & 5174 &    0.06$\pm$0.05     &                  &    0.53$\pm$0.08 & 3 &                 &     & $<$28.53       &   $<$--4.69     & 19\\ 
\object{HII  2880} & 5117 &    0.06$\pm$0.05     &                  &    0.68$\pm$0.08 & 3 &                 &     & 28.98$\pm$0.11 & --4.17$\pm$0.11 & 20\\   
\object{HII  3179} & 6172 &    0.06$\pm$0.05     &                  &    0.01$\pm$0.08 & 3 &                 &     & 29.27$\pm$0.18 & --4.49$\pm$0.18 & 20\\         
\multicolumn{2}{l}{\bf Hyades stars} &&&&&&&&&\\
\object{VB  15}    & 5772 &    0.12$\pm$0.03     &                  &    0.18$\pm$0.12 & 5 & --4.42$\pm$0.08 & 10  & 29.11$\pm$0.09 & --4.48$\pm$0.09 & 22  \\
\object{VB  19}    & 6300 &    0.12$\pm$0.03     &                  &    0.41$\pm$0.15 & 6 & --4.54$\pm$0.06 & 10  & 29.29$\pm$0.09 & --4.69$\pm$0.09 & 22  \\
\object{VB  25}    & 4790 &    0.12$\pm$0.03     &    0.04$\pm$0.08 &                  & 5 & --4.49$\pm$0.03 & 10  & $<$28.43       &   $<$--4.67     & 22  \\
\object{VB  31}    & 6045 &    0.12$\pm$0.03     &                  &    0.45$\pm$0.15 & 6 & --4.45$\pm$0.10 & 10  & 29.09$\pm$0.09 & --4.76$\pm$0.09 & 22  \\  
\object{VB  48}    & 6245 &    0.12$\pm$0.03     &                  &    0.39$\pm$0.15 & 6 & --4.50$\pm$0.05 & 10  & 29.28$\pm$0.07 & --4.64$\pm$0.07 & 22  \\ 
\object{VB  65}    & 6200 &    0.12$\pm$0.03     &                  &    0.34$\pm$0.15 & 6 & --4.58$\pm$0.07 & 10  & 29.06$\pm$0.08 & --4.80$\pm$0.08 & 22  \\
\object{VB  66}    & 6080 &    0.12$\pm$0.03     &                  &    0.18$\pm$0.15 & 6 & --4.46$\pm$0.08 & 10  & 29.30$\pm$0.07 & --4.56$\pm$0.07 & 22  \\
\object{VB  73}    & 5914 &    0.12$\pm$0.03     &                  &    0.07$\pm$0.10 & 5 & --4.50$\pm$0.08 & 10  & 29.06$\pm$0.09 & --4.63$\pm$0.09 & 22  \\
\object{VB  79}    & 5190 &    0.12$\pm$0.03     &    0.02$\pm$0.08 &                  & 5 & --4.44$\pm$0.06 & 10  & 28.63$\pm$0.16 & --4.71$\pm$0.16 & 22  \\
\object{VB  88}    & 6175 &    0.12$\pm$0.03     &                  &    0.33$\pm$0.15 & 6 & --4.56$\pm$0.09 & 10  &                &                 &   \\
\object{VB  97}    & 5887 &    0.12$\pm$0.03     &                  &    0.02$\pm$0.10 & 5 & --4.45$\pm$0.06 & 10  & 29.00$\pm$0.13 & --4.76$\pm$0.13 & 22  \\
\object{VB 105}    & 6055 &    0.12$\pm$0.03     &                  &    0.22$\pm$0.15 & 6 & --4.52$\pm$0.11 & 10  & 29.15$\pm$0.08 & --4.52$\pm$0.08 & 22  \\\hline
\end{tabular}
\end{table*}

\addtocounter{table}{-1}
\begin{table*}
\centering
\caption{Continued.}
\label{tab_data}
%\hspace*{0.4cm}
\scriptsize
\begin{tabular}{lcrrrcccccc} \hline\hline
ID$^a$    &    $T_{\rm eff}$ & \multicolumn{1}{c}{[Fe/H]$^b$}   &     \multicolumn{1}{c}{[O/Fe] (6300)}  &     \multicolumn{1}{c}{[O/Fe] (Triplet)}
 & Ref. & log($R_{\rm HK}^{\prime}$)$^c$ & Ref. & log($L_{\rm X}$)$^d$ &  log($R_{\rm X}$)$^d$  &  Ref.\\
   &  (K) &  &  &  &  & &  & (ergs s$^{-1}$) &    &  \\\hline
\multicolumn{2}{l}{\bf Field FG dwarfs} &&&&&&&&&\\
\object{HD   3079} & 5892 &  --0.19$\pm$0.07 &  --0.05$\pm$0.16 &    0.39$\pm$0.16 & 7 & --4.99          & 11    &                  &                 &   \\
\object{HD   4614} & 5900 &  --0.27$\pm$0.10 &    0.08$\pm$0.09 &    0.08$\pm$0.11 & 8 & --4.93          & 11    &   27.41$\pm$0.07 & --6.26$\pm$0.07 &  23 \\ 
\object{HD   9562} & 5930 &    0.20$\pm$0.06 &  --0.11$\pm$0.06 &  --0.03$\pm$0.08 & 9 & --5.19          & 11    &                  &                 &   \\
\object{HD  10307} & 5883 &    0.02$\pm$0.10 &  --0.02$\pm$0.07 &  --0.22$\pm$0.12 & 8 & --5.02          & 12    &  $<$27.64        & $<$--6.13       & 24\\
\object{HD  14412} & 5350 &  --0.47$\pm$0.06 &    0.12$\pm$0.06 &    0.15$\pm$0.08 & 9 & --4.85          & 11    &  $<$27.08        & $<$--6.13       & 25     \\
\object{HD  17051} & 6150 &    0.14$\pm$0.06 &  --0.15$\pm$0.06 &    0.11$\pm$0.08 & 9 & --4.65          & 13    &   28.83$\pm$0.07 & --4.97$\pm$0.07 &  23 \\
\object{HD  18262} & 6454 &    0.20$\pm$0.10 &    0.07$\pm$0.13 &    0.12$\pm$0.06 & 8 & --5.04          & 11    &                  &                 &   \\
\object{HD  19994} & 6240 &    0.19$\pm$0.06 &  --0.13$\pm$0.06 &    0.04$\pm$0.08 & 9 & --4.88          & 11    &                  &                 &   \\
\object{HD  20630} & 5673 &    0.03$\pm$0.10 &    0.01$\pm$0.15 &  --0.11$\pm$0.11 & 8 & --4.45          & 12    &   28.89$\pm$0.04 & --4.63$\pm$0.04 &  23 \\
\object{HD  20807} & 5821 &  --0.21$\pm$0.10 &  --0.10$\pm$0.12 &  \multicolumn{1}{c}{--0.10}  & 8 & --4.79          & 13    &  $<$27.14        & $<$--6.43       &25\\ 
\object{HD  22484} & 5983 &  --0.10$\pm$0.10 &  --0.02$\pm$0.08 &    \multicolumn{1}{c}{0.04}  & 8 & --5.12          & 11    &   27.19$\pm$0.09 & --6.87$\pm$0.09 &24\\
\object{HD  23249} & 5020 &    0.24$\pm$0.06 &  --0.16$\pm$0.06 &  --0.11$\pm$0.08 & 9 & --5.22          & 13    &   26.95$\pm$0.18 & --7.14$\pm$0.18 &  23 \\ 
\object{HD  26491} & 5745 &  --0.18$\pm$0.10 &  --0.01$\pm$0.08 &  \multicolumn{1}{c}{--0.06}  & 8 & --4.95          & 13    &                  &                 &   \\
\object{HD  30562} & 5926 &    0.14$\pm$0.10 &  --0.10$\pm$0.08 &  \multicolumn{1}{c}{--0.19} & 8 & --5.04          & 11    &   &  &   \\ 
\object{HD  34411} & 5852 &    0.03$\pm$0.10 &    0.01$\pm$0.09 &  --0.12$\pm$0.10 & 8 & --5.07          & 12    &  $<$27.40        & $<$--6.44       &24\\
\object{HD  42618} & 5653 &  --0.16$\pm$0.07 &    0.12$\pm$0.16 &    0.30$\pm$0.16 & 7 & --4.94          & 11    &                  &                 &   \\
\object{HD  43834} & 5550 &    0.10$\pm$0.06 &  --0.13$\pm$0.06 &    0.10$\pm$0.08 & 9 & --4.94          & 13    &   27.46$\pm$0.13 & --6.05$\pm$0.13 &  23 \\ 
\object{HD  45184} & 5820 &    0.04$\pm$0.06 &  --0.07$\pm$0.06 &    0.11$\pm$0.08 & 9 & --4.95          & 11    &                  &                 &   \\
\object{HD  48938} & 6010 &  --0.37$\pm$0.10 &  --0.01$\pm$0.09 &  \multicolumn{1}{c}{--0.07} & 8 & --4.96          & 11    &                  &                 &   \\
\object{HD  71148} & 5703 &  --0.08$\pm$0.07 &    0.05$\pm$0.16 &    0.27$\pm$0.16 & 7 & --4.95          & 11    &                  &                 &   \\
\object{HD 102870} & 6149 &    0.13$\pm$0.10 &  --0.05$\pm$0.08 &  --0.04$\pm$0.07 & 8 & --4.94          & 11    &   28.39$\pm$0.04 & --5.74$\pm$0.04 &  23 \\ 
\object{HD 108309} & 5710 &    0.10$\pm$0.10 &  --0.03$\pm$0.09 &  \multicolumn{1}{c}{--0.16}  & 8 & --4.95          & 13    &                  &                 &   \\
\object{HD 114710} & 6009 &    0.06$\pm$0.10 &  --0.02$\pm$0.08 &  --0.16$\pm$0.07 & 8 & --4.76          & 11    &   28.06$\pm$0.06 & --5.66$\pm$0.06 &  23 \\ 
\object{HD 121560} & 6081 &  --0.38$\pm$0.07 &    0.17$\pm$0.16 &    0.37$\pm$0.16 & 7 & --4.92          & 11    &                  &                 &   \\
\object{HD 141004} & 5915 &  --0.03$\pm$0.10 &    0.16$\pm$0.05 &  --0.05$\pm$0.08 & 8 & --4.97          & 11    &   27.67$\pm$0.10 & --6.20$\pm$0.10 &  23 \\ 
\object{HD 142860} & 6276 &  --0.14$\pm$0.10 &    0.13$\pm$0.10 &  --0.01$\pm$0.08 & 8 & --4.82          & 11    &   27.28$\pm$0.15 & --6.76$\pm$0.15 &  23 \\ 
\object{HD 144585} & 5880 &    0.33$\pm$0.06 &  --0.21$\pm$0.06 &  --0.05$\pm$0.08 & 9 & --5.10          & 11    &                  &                 &   \\
\object{HD 147513} & 5880 &    0.03$\pm$0.06 &  --0.05$\pm$0.06 &  --0.02$\pm$0.08 & 9 & --4.52          & 13    &   28.96$\pm$0.03 & --4.61$\pm$0.03 &  23 \\ 
\object{HD 147584} & 6090 &  --0.06$\pm$0.06 &  --0.04$\pm$0.06 &    0.12$\pm$0.08 & 9 & --4.56          & 13    &   29.11$\pm$0.04 & --4.59$\pm$0.04 &  23 \\ 
\object{HD 154417} & 6167 &    0.09$\pm$0.06 &  --0.07$\pm$0.06 &    0.00$\pm$0.08 & 9 & --4.59          & 11    &   28.82$\pm$0.05 & --4.90$\pm$0.05 &  23 \\ 
\object{HD 156365} & 5820 &    0.23$\pm$0.06 &  --0.13$\pm$0.06 &    0.10$\pm$0.08 & 9 & --5.17          & 11    &                  &                 &   \\
\object{HD 157347} & 5720 &    0.03$\pm$0.06 &  --0.08$\pm$0.06 &  --0.03$\pm$0.08 & 9 & --5.04          & 11    &                  &                 &   \\
\object{HD 157466} & 6050 &  --0.39$\pm$0.06 &    0.08$\pm$0.06 &    0.17$\pm$0.08 & 9 & --4.89          & 11    &                  &                 &   \\
\object{HD 160691} & 5800 &    0.32$\pm$0.06 &  --0.23$\pm$0.06 &  --0.03$\pm$0.08 & 9 & --5.02          & 13    &   27.12$\pm$0.06 & --6.72$\pm$0.06 &25\\
\object{HD 172051} & 5580 &  --0.24$\pm$0.06 &    0.01$\pm$0.06 &    0.06$\pm$0.08 & 9 & --4.90          & 11    &   27.63$\pm$0.15 & --5.77$\pm$0.15 &  23 \\
\object{HD 179949} & 6200 &    0.16$\pm$0.06 &  --0.15$\pm$0.06 &    0.12$\pm$0.08 & 9 & --4.79          & 11    &   28.61$\pm$0.11 & --5.24$\pm$0.11 &  23 \\ 
\object{HD 182572} & 5600 &    0.37$\pm$0.06 &  --0.22$\pm$0.06 &    0.06$\pm$0.08 & 9 & --5.10          & 11    &   27.59$\pm$0.14 & --6.26$\pm$0.14 &  23 \\
\object{HD 185144} & 5237 &  --0.24$\pm$0.10 &  --0.07$\pm$0.16 &  --0.04$\pm$0.13 & 8 & --4.85          & 11    &   27.61$\pm$0.02 & --5.59$\pm$0.02 &  23 \\ 
\object{HD 186408} & 5790 &    0.08$\pm$0.10 &  --0.22$\pm$0.08 &  --0.17$\pm$0.09 & 8 & --5.10          & 11    &                  &                 &   \\
\object{HD 186427} & 5719 &    0.03$\pm$0.10 &  --0.09$\pm$0.08 &  --0.17$\pm$0.10 & 8 & --5.08          & 11    &                  &                 &   \\
\object{HD 190248} & 5585 &    0.37$\pm$0.06 &  --0.23$\pm$0.06 &    0.00$\pm$0.08 & 9 & --5.00          & 13    &   27.26$\pm$0.17 & --6.44$\pm$0.17 &  23 \\ 
\object{HD 190406} & 5880 &  --0.11$\pm$0.10 &    0.15$\pm$0.15 &    0.05$\pm$0.08 & 8 & --4.77          & 11    &   28.45$\pm$0.12 & --5.23$\pm$0.12 &  23 \\ 
\object{HD 193307} & 5960 &  --0.32$\pm$0.06 &    0.09$\pm$0.06 &    0.34$\pm$0.08 & 9 & --4.90          & 13    &                  &                 &   \\
\object{HD 197214} & 5570 &  --0.22$\pm$0.06 &    0.05$\pm$0.06 &    0.12$\pm$0.08 & 9 & --4.92          & 11    &   27.86$\pm$0.18 & --5.59$\pm$0.18 &  26 \\
\object{HD 199960} & 5940 &    0.27$\pm$0.06 &  --0.18$\pm$0.06 &  --0.01$\pm$0.08 & 9 & --5.08          & 11    &   27.76$\pm$0.18 & --6.12$\pm$0.18 &  23 \\ 
\object{HD 206860} & 5926 &    0.03$\pm$0.10 &  --0.04$\pm$0.10 &  --0.10$\pm$0.09 & 8 & --4.42          & 13    &   29.25$\pm$0.04 & --4.41$\pm$0.04 &  23 \\
\object{HD 210277} & 5500 &    0.22$\pm$0.06 &  --0.11$\pm$0.06 &    0.07$\pm$0.08 & 9 & --5.06          & 11    &                  &                 &   \\
\object{HD 217014} & 5789 &    0.20$\pm$0.06 &  --0.15$\pm$0.06 &  --0.04$\pm$0.08 & 9 & --5.08          & 11    &                  &                 &   \\
\object{HD 217107} & 5620 &    0.35$\pm$0.06 &  --0.24$\pm$0.06 &  --0.03$\pm$0.08 & 9 & --5.08          & 11    &                  &                 &   \\
\object{HD 217877} & 5872 &  --0.18$\pm$0.07 &    0.00$\pm$0.16 &    0.34$\pm$0.16 & 7 & --5.01          & 11    &                  &                 &   \\
\object{HD 222368} & 6157 &  --0.21$\pm$0.10 &    0.13$\pm$0.09 &    0.21$\pm$0.05 & 8 & --4.95          & 12    &   27.97$\pm$0.08 & --6.15$\pm$0.08 &  23 \\
\object{HD 224022} & 6100 &    0.12$\pm$0.06 &  --0.12$\pm$0.06 &    0.09$\pm$0.08 & 9 & --4.97          & 13    &                  &                 &   \\\hline 
\end{tabular}
\end{table*}

\begin{table*}
\normalsize
\begin{tabular}{lcrrrcccccc} 
\end{tabular}
Key to references: [1] M04; [2] M03; [3] Schuler \etal (2004); [4] King \etal (2000); [5] King \& Hiltgen (1996); [6] Garc\'{\i}a L{\'o}pez \etal (1993); [7] Reddy \etal (2003); [8] King \& Boesgaard (1995); [9] Bensby \etal (2004); [10] Paulson \etal (2002); [11] Wright \etal (2004); [12] Soderblom (1985); [13] Henry \etal (1996); [14] Dempsey \etal (1993); [15] Dempsey \etal (1997); [16] Schwope \etal (2000); [17] Neuh\"auser \etal (2000); [18] H\"unsch \etal (1998a); [19] Micela \etal (1999); [20] Stauffer \etal (1994); [21] Micela \etal (1996); [22] Stern, Schmitt, \& Kahabka (1995); [23] H\"unsch, Schmitt, \& Voges (1998b); [24] Schmitt \& Liefke (2003); [25] Schmitt (1997); [26] H\"unsch \etal (1999).\\ 
$^a$ Identification for Pleiades and Hyades stars from Hertzsprung (1947) and van Bueren (1952), respectively.\\
$^b$ We quote the mean cluster metallicity of the Pleiades and Hyades determined by King \etal (2000) and Cayrel, Cayrel de Strobel, \& Campbell (1985), respectively.\\
$^c$ For the RS CVn binaries and field subgiants, a blank indicates that the \cas profiles were not filled in by emission. The errors take into account the uncertainties in $(V-R)_0$ and $T_{\rm eff}$.\\
$^d$ Except for the open cluster stars and \object{HD 118238} (130 pc; Strassmeier \etal 1993), all luminosities were rescaled to the {\em Hipparcos} distances. For consistency with previous X-ray observations, the bolometric luminosities were computed adopting a mean distance to the Pleiades of 127 pc (Crawford \& Perry 1976). For the Hyades, we used the individual distances of Schwan (1991). No reddening and a mean extinction $A_V$=0.12 mag were assumed for the Hyades and Pleiades, respectively (Nissen 1988). We used $A_V$=0.26 mag for HII 676 (Stauffer \etal 1994).  The bolometric corrections are taken from Flower (1996). The errors take into account the uncertainties in the X-ray counts and in the distances.\\  
\end{table*}

\end{document}